\documentclass[mathleft
]{an}
\usepackage{graphicx}
\usepackage{times}
\begin{document}

% The following seven commands are intended for editorial usage and should be ignored by
% the author(s).
\Pagespan{789}{}% Document's page range. 
% If second parameter is left empty, the last page is computed automatically.
\Yearpublication{2011}%
\Yearsubmission{2011}%
\Month{11}%   
\Volume{999}%  
\Issue{88}% 
% \DOI{This.is/not.aDOI}% 

\title{Horizon synthesis for archaeo-astronomical purposes}

\author{F. Patat\inst{1}\fnmsep\thanks{Corresponding author:
  \email{fpatat@eso.org}\newline}
%Example 
%for footnote, note the usage of the \texttt{fnmsep}
%command as separator between institute number and footnote mark} 
%\and  G.H. Ostwriter\inst{2,3}
}

\titlerunning{Horizon synthesis for archaeo-astronomical purposes}

\authorrunning{F. Patat}

\institute{
European Southern Observatory - 
     K.-Schwarzschild-Str. 2, D-85748 Garching b. M\"unchen, Germany
}

\received{17 March 2011}
\accepted{4 July 2011}
\publonline{later}


\abstract{In this paper I describe a simple numerical procedure to
  compute synthetic horizon altitude profiles for any given site. The
  method makes use of a simplified model of local Earth's curvature,
  and it is based on the availability of digital elevation models
  describing the topography of the area surrounding the site under
  study. Examples constructed using the Shuttle Radar Topographic
  Mission (SRTM) data (with 90m horizontal resolution) are
  illustrated, and compared to direct theodolite measurements. The
  proposed method appears to be reliable and applicable in all cases
  when the distance to the local horizon is larger than $\sim$10 km,
  yielding a rms accuracy of $\sim$0.1 degrees (both in azimuth
    and elevation).  Higher accuracies can be achieved with higher
  resolution digital elevation models, like those produced by many
  modern national geodetic surveys.}

\maketitle

\section{\label{sec:intro} Introduction}

When studying the orientation of buildings, tombs, or any other man
made structure, having an handy description of the natural horizon is
a fundamental step. In most of the cases this is done directly
measuring the horizon altitude at the relevant azimuths using, for
instance, a theodolite and sun fixes. Depending on the angular
sampling one wants to achieve, this might turn into a rather long and
boring procedure. An alternative possibility is the calibration of a
few points using direct theodolite readings coupled to digital,
rectified photography. However, this might turn to be difficult, or
even impossible, in the case the natural horizon is not visible
because of modern constructions or vegetation.

In this paper I propose an alternative solution, which is based on a
simple geometrical model for Earth's local shape and the availability
of a digital elevation model (DEM) for the area under examination. The
idea is rather simple. For a given observer's site, the line of site
(hereafter LOS) elevation profile (LOSEP) along the input azimuth is
extracted. Then, for each point along the LOS, the apparent altitude
above the local horizontal is estimated and the maximum is found. By
definition, this is the natural horizon altitude as seen from the
observing site. Repeating the same procedure for all azimuths (with a
given angular step) will finally allow one to retrieve the horizon
altitude profile. The angular resolution clearly depends on the
horizontal sampling in the DEM, while the accuracy is related to the
DEM vertical accuracy.

Since the available DEMs provide the elevation above a given ellipsoid
(typically the WGS84), the first thing one needs to take into account
is Earth's curvature, which makes a far mountain appear lower than it
would if the Earth's surface were flat. Then, although this is a
second order effect, one has to correct for the terrestrial refraction
(which has the opposite effect). These two corrections are discussed
in the next two sections.

\section{\label{sec:curv} Correcting for Earth's curvature}

Let us consider an observer placed in $A$ at an elevation $h_A$ (above
the sea level), and a point $B$ (elevation $h_B$), located at a
distance $d$ from A.  If Earth were flat, the altitude $\alpha$ of $B$
seen from $A$ would simply be:

\begin{displaymath}
\alpha=\arctan \left( \frac{h_B-h_A}{d} \right).
\end{displaymath}

However, because of Earth's curvature, far objects subtend altitudes
which are smaller than those given by the previous expression. Using a
simplified model for Earth's local curvature, one can derive the
following approximate expression (see Appendix~\ref{sec:app}):

\begin{equation}
\label{eq:curv}
\alpha \approx \arctan \left [
\frac{h_B-h_A}{d} - \frac{1}{2} \frac{d}{R}
\right ],
\end{equation}

\noindent where $R$ is the local radius of curvature (see
Equation~\ref{eq:radius}).  For example, suppose that a $h_B$=3.0 km
mountain peak is observed from a site placed $d$=100 km away, 
at an elevation $h_A$=0.1 km.  Using
$R$=6370 km one obtains $\alpha$=1$^\circ$.21. Neglecting Earth's
curvature one would get $\alpha$=1$^\circ$.66, almost half a degree
off.  It can be easily shown that for ellipsoidal distances $d\leq$300
km the error introduced by the approximated expression (\ref{eq:curv})
is less than 0.01 degrees, which is well below the typical
archaeoastronomical needs.

In the cases where $|h_B-h_A|/d\ll 1$, using Equation~\ref{eq:curv}
one can show that the apparent altitude decreases at a rate of $1/2R$,
which is about 16$^{\prime\prime}$.2 km$^{-1}$ (or 0$^\circ$.0045
km$^{-1}$).

\begin{figure}
\centering
\includegraphics[width=8cm]{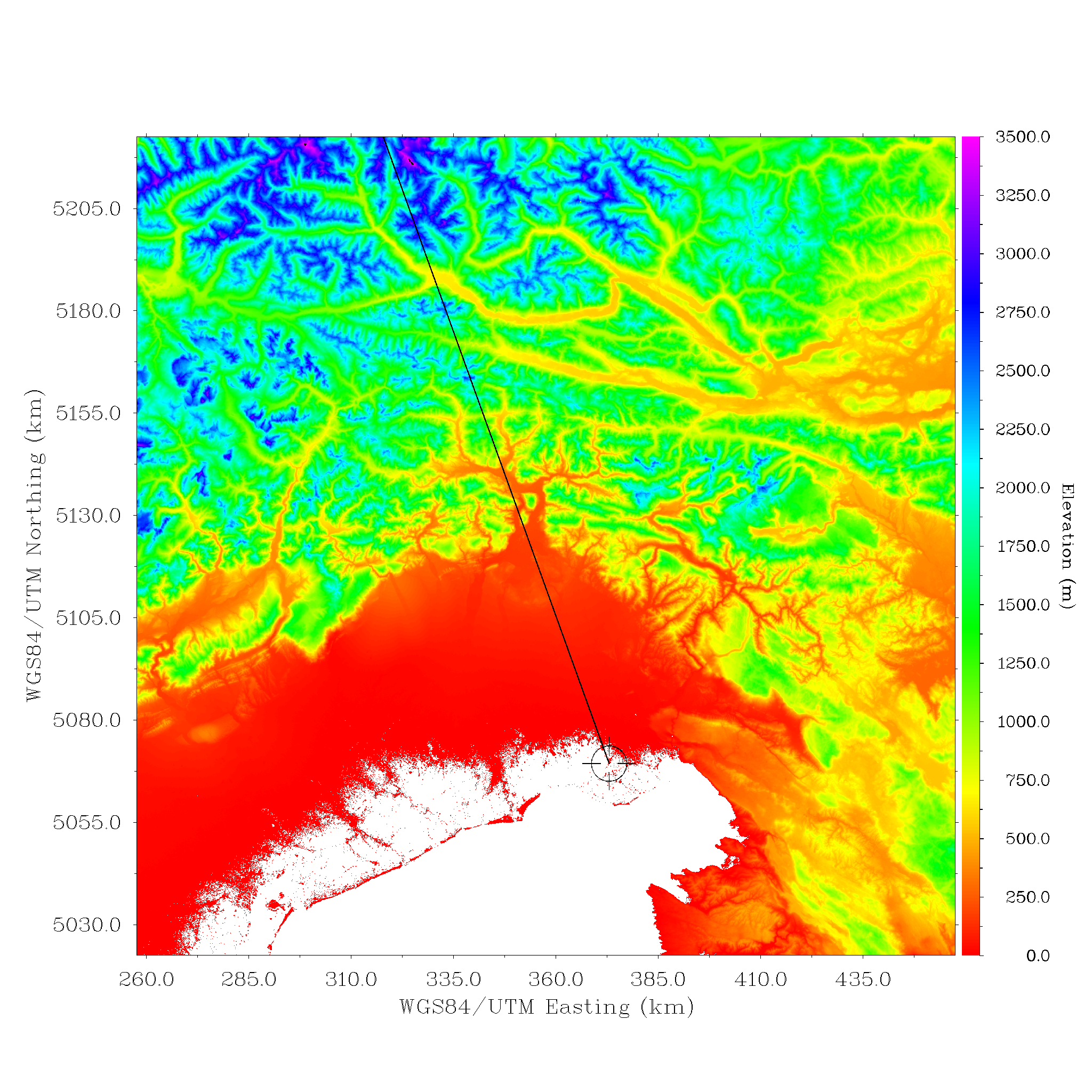}  
\caption{\label{fig:losep}Example LOSEP extraction for the Roman town
  Aquileia ($\lambda$=13$^\circ$.367 E, $\phi$= 45$^\circ$.767 N),
  with $\varphi$=339$^\circ$. The underlying DEM is from the SRTM90
  data set, and is displayed in the Universal Transverse Mercator
  projection (UTM, 33N). The white area in the lower part of the map
  is the Adriatic Sea.}
\end{figure}

\section{\label{sec:refr}Correcting for terrestrial refraction}

As the light travels across Earth's atmosphere it bends, in such a way
that a far object appears to be higher than actually is. This
phenomenon is known as terrestrial refraction\footnote{As opposed to
  the so called astronomical refraction. The phenomenon is very similar,
  but in that case the light rays have to cross the whole atmosphere,
  which makes its description much more complex, especially when one
  is to consider objects very close to the horizon.}.

If $\alpha$ is the unrefracted horizon altitude (corrected for
Earth's curvature), then the apparent horizon altitude
$\alpha^\prime$ is simply given by $\alpha^\prime=\alpha+R_T$.
Several analytical descriptions of $R_T$ have been proposed, but the
one developed by Bomsford (\cite{bomsford}) appears to be the most
accurate (see for instance Sampson et al. \cite{sampson03}). In this
formulation the terrestrial refraction is given by

\begin{displaymath}
R_T = \kappa d \; ,
\end{displaymath}

\noindent where $d$ is the distance between the observer an the natural horizon
(in km), and $\kappa$ (degrees km$^{-1}$) is defined as:

\begin{equation}
\kappa = \frac{1}{R} \frac{0.252 P}{T^2} 
\left ( 34.2 + \frac{dT}{dz} \right ) \frac{180}{\pi} \; ,
\end{equation}

\noindent where $R$ is Earth's radius (in km), $P$ is the atmospheric
pressure (in millibars), $T$ is the ground air temperature (in K), and
$dT/dz$ is the ground atmospheric vertical temperature gradient (in K
km$^{-1}$). For typical atmospheric conditions ($P$=1000 mb, $T$=293
K, $dT/dz$=$-$10 K km$^{-1}$) one finds that $\kappa\sim$2.3
arcsec km$^{-1}$ (or 0.00064 degrees km$^{-1}$). This is about seven
times smaller than the effect produced by Earth's curvature. In the
case of the example discussed in the previous section ($d$=100 km)
this would turn into a correction of $\sim$0.06 degrees.

One can simplify the previous formula introducing the refraction
constant $K$:

\begin{displaymath}
 K=0.273 \left (34.2 + \frac{dT}{dz}  \right ) \;.
\end{displaymath}

With this setting, $\kappa$ (in degrees km$^{-1}$) can be written as:

\begin{equation}
\kappa = 0.00829 \; K \; \frac{P}{T^2}\; .
\end{equation}

Typical values of $K$ range from $\sim$5 (during the day) to $\sim$10
(at sunset/sunrise or night). The fluctuation of atmospheric
conditions introduce a variation in $\kappa$, which can range from
$\sim$1.5 to $\sim$4 arcsec km$^{-1}$. For instance, on a distance of
100 km this translates into a peak-to-peak horizon altitude change of
$\sim$0.07 degrees. Therefore, even in the hypothesis all other
quantities are known with negligible errors, this sets the {\em natural}
maximum accuracy one can achieve on the horizon apparent altitude.

\section{Digital Elevation Model \label{sec:dem}}

The proposed method is based on the availability of a DEM, i.e. an
array of values giving the elevation of a given point above the
underlying ellipsoid. If $\lambda$ and $\phi$ are the longitude and
latitude of a point $\vec{P}$ on the ellipsoid, I will express its
elevation $z$ as $z=g(\lambda,\phi)$ or, alternatively, as $z=g(\vec{P})$.

Because of its nature, a DEM is a discrete collection of data points,
obtained with some spatial resolution. Typically, the data are
distributed on a regular grid, with a horizontal sampling that I will
indicate as $\Delta l$. This corresponds to the minimum scale one can
resolve in the DEM. Obviously, if one is to get the elevation of a
point which does not coincide with one of the grid nodes, then one
will have to use some interpolation method (nearest neighbor, linear
interpolation, bi-cubic spline interpolations, etc). No matter what
method is used, though, the resolution is dictated by the sampling.
In the following I will assume that the DEM is given with a regular
sampling $\Delta l$, which is to say that the data points have been
obtained/re-gridded at a constant step. Also, I will indicate with
$\sigma_z$ the rms uncertainty on each DEM data point.

Each country has its own geographical survey program-me, aiming at
mapping its territory with a certain resolution. These data may or may
not be available to the reader. Therefore, in this article I will
consider the Shuttle Radar Topographic Mission, which has the
advantage of having a relatively good horizontal resolution, a good
vertical accuracy, an almost world wide coverage, and, most
importantly, is freely available\footnote{The data can
  be freely downloaded at the following URL:
  http://srtm.csi.cgiar.org/SELECTION/inputCoord.asp}.

\subsection{\label{sec:srtm} The Shuttle Radar Topographic Mission}

The Shuttle Radar Topographic Mission (SRTM) provides an homogeneous
coverage of Earth's elevation (Farr et al. \cite{srtm}). For the US
territory it is distributed with a horizontal resolution of 30 m
(SRTM30), while for the rest of the planet data were averaged within
bins of 90$\times$90 m$^2$ (SRTM90), hence including 3$\times$3
original data points each. The 90\%-level absolute vertical accuracy
of the SRTM original data is better than 9 m (Farr et
al. \cite{srtm}), which corresponds to an rms accuracy of 5.5
m. Therefore, the formal rms error on the 90 m data is expected to be
$\sigma_z\sim$1.8 m. I note, however, that this is strictly true only
if the terrain is smooth on scales smaller than 30 m, otherwise the
averaging process within 90 m bins produces an artificial smoothing,
turning into much larger deviations from the real values. Clearly this
is going to affect very steep mountain regions, where the elevation
can change significantly on scales of tens of meters. As a
consequence, sharp mountain peaks appear in the SRTM90 data with lower
than real elevations.

The altitude uncertainty $\sigma_\alpha$ (in degrees) implied by an
elevation uncertainty $\sigma_z$ can be estimated through the
following approximate expression:

\begin{displaymath}
\sigma_\alpha \approx \frac{180\sqrt{2}}{\pi} \; \frac{\sigma_z}{d}\;,
%\sqrt{\frac{1}{1+\Delta^2z/d^2}}
\end{displaymath}

\noindent where $d$ is the distance between the observer and the point
being surveyed, expressed in the same units as $\sigma_z$. For
  instance, the formal elevation error of the SRTM90 data (1.8 m)
  translates into an altitude error of about 0.1 degrees at
  $d\simeq$1.5 km.

As for the SRTM horizontal accuracy, this is better than 20 m (at the
90\%-level; Farr et al. \cite{srtm}). This corresponds to an rms
deviation $\sigma_l$=14 m). Therefore, in the SRTM90 case, this is
much better than the horizontal resolution $\Delta l$. As a
consequence, the rms error on the azimuth $\varphi$ of a given
data point is approximately

\begin{displaymath}
\sigma_\varphi \approx \frac{180}{\pi} \; \frac{\sigma_l}{d} 
\simeq \frac{0.8}{d_{km}} \;.
\end{displaymath}

At $d$=10 km this corresponds to an uncertainty of $\sim$0.1 degrees,
which is a fifth of the apparent diameter of the sun (and the
moon). Obviously, if the horizon is farther away, the azimuth
uncertainty becomes proportionally smaller. If one is after higher
azimuth accuracies, or the horizon in the direction of interest is
closer than 10 km, then a DEM with better sampling is
required\footnote{In many countries surveys with horizontal
  resolutions of 15 m or better are either available or under
  construction. The reader should get in touch with the local
  authorities to verify the public availability of higher resolution
  data.}.

\begin{figure}
\centering
\includegraphics[width=8cm]{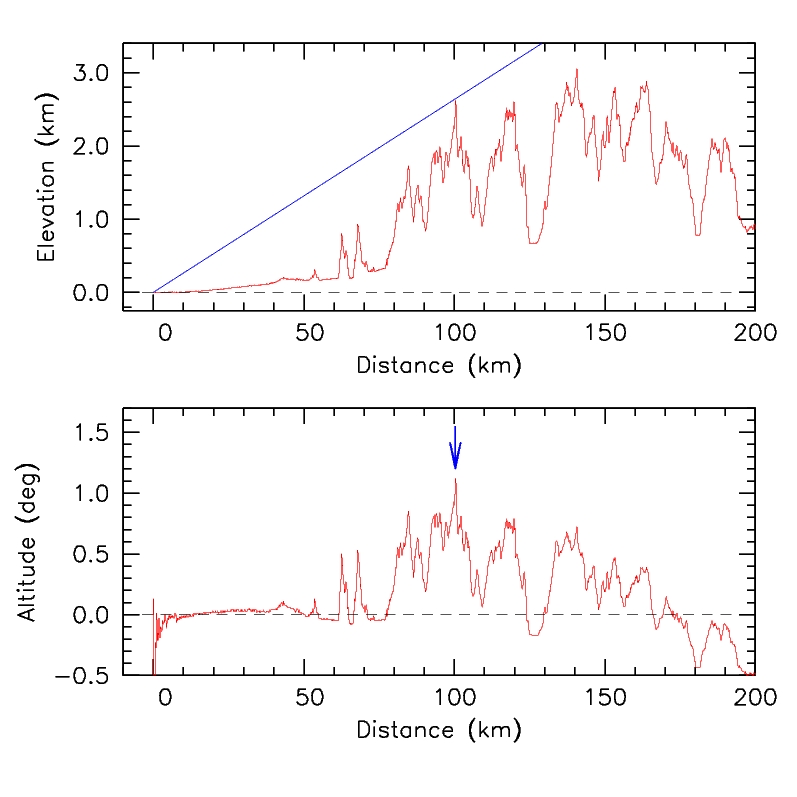}  
\caption{\label{fig:aq}Upper panel: LOSEP along $\varphi$=339$^\circ$
  for the Roman town of Aquileia (see Figure~\ref{fig:losep}). Lower
  panel: the corresponding altitude profile. The blue arrow marks the
  maximum altitude, corresponding to the natural horizon along the
  chosen direction.}
\end{figure}

In this respect, I note that a preliminary version of the data
produced by the Advanced Spaceborne Thermal Emission and Reflection
Radiometer (ASTER\footnote{See:
  http://asterweb.jpl.nasa.gov/index.asp}) was recently
released. Nominally they have an horizontal resolution of 30 m, which
is a factor 3 better than the SRTM90. However, a close look to the
data shows that the effective horizontal resolution is actually close
to 90 m and the vertical accuracy is not as good as the one of SRTM,
at least not in the released version.

\section{\label{sec:losep}LOSEP extraction and horizon determination}

Having a DEM at hand, one can extract the LOSEP starting from the
observer's position $\vec{P_0}$ along a given azimuth $\varphi$ as a
function of distance $d$ from the observer. This corresponds to
solving the forward geodesic problem, i.e. computing the running end
point $\vec{P}$ of a geodesic path on the ellipsoid, given the start
point $\vec{P_0}$, a path length $d$, and a starting azimuth
$\varphi$. As is standard in geodesy, this is done numerically, using
the iterative algorithm devised by Vincenty
(\cite{vincenty}). Although the solution is non-analytical, for the
sake of clarity I will indicate it as

\begin{displaymath}
\vec{P}=f(\vec{P_0},d,\varphi)\;.
\end{displaymath}

Given the discrete nature of the DEM, the LOSEP is extracted at a
number of points, which are separated by some constant
length. Obviously, the maximum resolution is attained when this length
is equal to the DEM resolution $\Delta l$.

\begin{figure*}
\centering
\includegraphics[width=8cm]{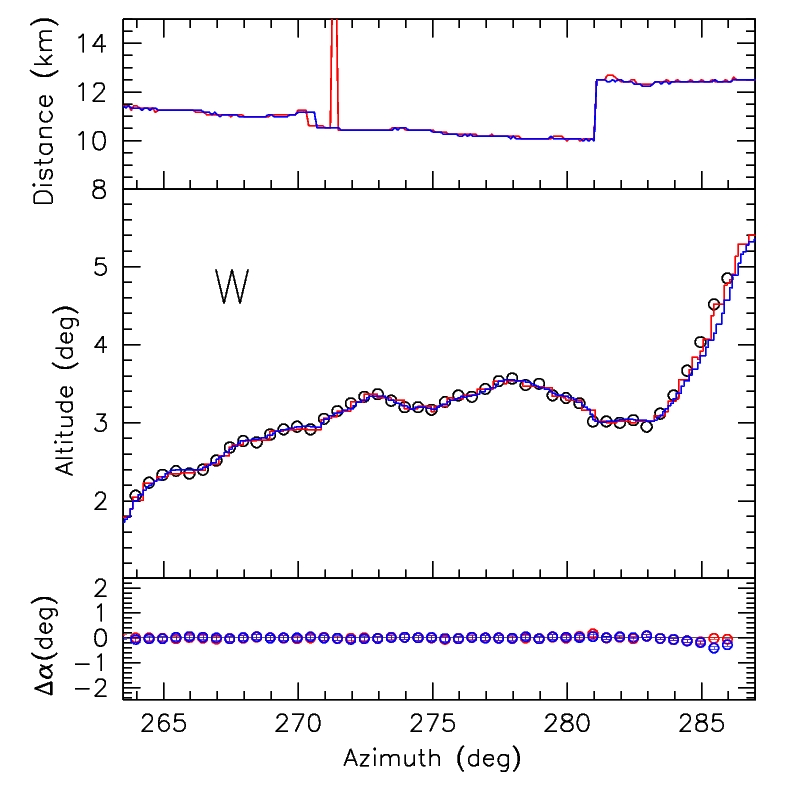} 
\includegraphics[width=8cm]{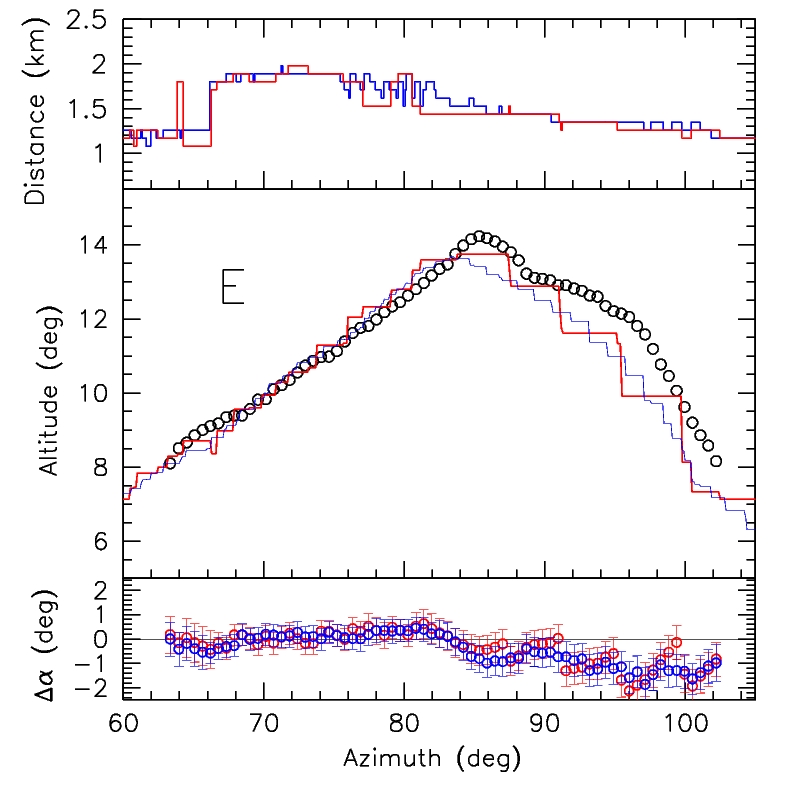} 
\caption{\label{fig:comp} Comparison between the theodolite readings
  (empty circles) and the synthetic profiles (W to the left and E to
  the right) computed using SRTM (red) and ASTER (blue) DEMs for the
  site of St. Martin (Artegna - Italy). Top panel: horizon
    distance. Middle panel: horizon profiles. Bottom panel: $\Delta
    \alpha$ residuals (computed$-$measured). The errorbars indicate
    the 5-$\sigma$ error of the synthetic profile.}
\end{figure*}

With these definitions, the LOSEP extraction is done through the
following steps:

\begin{enumerate}
\item elevation $z_0=g(\vec{P_0})$ is extracted from the DEM at
      ($\lambda_0, \phi_0$);
\item \label{item:two} the path length for the $i$-th step along the 
       LOS is updated to $d_i=\Delta l \; i$;
\item the coordinates ($\lambda_i,\phi_i$) of the running point $\vec{P}$ are
  computed as $f(\vec{P_0},d_i,\varphi)$
\item the elevation of the running point is computed as 
      $z_i=g(\lambda_i,\phi_i)$;
\item the apparent altitude $\alpha_i$ of $\vec{P}$ as seen from
  $\vec{P_0}$ is calculated with Equation~\ref{eq:approx} and corrected
  for terrestrial refraction;
\item the cycle is repeated from step \ref{item:two}, until $d_i$ reaches
     a maximum value $d_{max}$.
\end{enumerate}

Once this is done, the horizon elevation along azimuth $\varphi$ is
determined as $\alpha(\varphi)=\mbox{max}\{\alpha_i\}$. I note that
in most of the cases it is sufficient to consider distances within
$d_{max}\simeq$200 km. The exact value depends on the maximum
elevation in the area surrounding the site under study. From
Equation~\ref{eq:curv}, one has that the distance at which a
mountain peak appears at an altitude $\alpha$=0 is given by
$d\approx\sqrt{2R\Delta z}$, where $\Delta z$ is the difference in
elevation. For $\Delta z\leq$4.0 km (which is true for most of the
sites on the planet), it is $d\leq$225 km.  Incidentally, here emerges
one of the advantages of horizon synthesis over direct
measurements. If the horizon is far away ($d>$50-100 km), it might be
very difficult, if not impossible, to have a sufficient atmospheric
transparency to be able to actually see it.

An example LOSEP extraction is presented in Figure~\ref{fig:losep},
where I have chosen the Roman town Aquileia (Italy; founded in 181
BCE) as the observing site. For the sake of the example I have traced
the LOS along one of the two cardinal directions of the town
($\varphi\simeq$339$^\circ$), in the NW direction. The site is marked
with a circle, while the selected line of sight is traced by a solid
line. The underlying DEM (from the SRTM90 data set) has been projected
to the Universal Transverse Mercator system (see Hager, Behensky \&
Drew \cite{utm}).

The LOSEP extracted along this direction is shown in
Figure~\ref{fig:aq} (upper panel), and it reaches about 3000 m in
the Austrian Alps, at a distance of 140 km from Aquileia. However, the
horizon is located at about 100 km, at an elevation of $\sim$2600 m,
and it subtends an apparent altitude of $\sim$1$^\circ$.1
(Figure~\ref{fig:aq}, lower panel).

Clearly the whole process can be repeated for $\varphi$ ranging from
0$^\circ$ to 360$^\circ$ with a given step $\Delta \varphi$. This will
finally give the full horizon profile $\alpha(\varphi)$.

\section{\label{sec:results}Validation}

The simplest way of validating the procedure outlined in the previous
section is a comparison between the synthetic profile and direct
horizon measurements. The natural horizon altitude can be determined
using a theodolite and sun fixes, taking measurements with some
azimuth step. The typical accuracy one can achieve with this method is
of the order of a tenth of a degree or better, depending on the
quality of the instrument used\footnote{It must be noticed that
  when the horizon is close, trees can substantially affect direct
  measurements, producing systematically higher theodolite readings.}. 

In the course of studying the orientation of the ancient church of
St. Martin (Artegna - Italy; 13$^\circ$.1528 E, 46$^\circ$.2415 N, 267
m a.s.l.), I had measured the natural horizon altitude in a range
around E and W directions. The building is located on top of
St. Martin Hill, about 50 m above the surrounding plain. The E horizon
is dominated by the presence of Mt. Faet (750 m a.s.l.), whose top is
located at less than 2 km from St. Martin. Seen from the site, the top
subtends an angle of about 14 degrees. On the contrary, the W§ horizon
is located at more than 10 km, and has an altitude around 3 degrees.
The different distances to the E and W horizon locations from the
observing site make this an ideal test case. The comparison between
the theodolite readings and the synthetic profile is presented in
Fig.~\ref{fig:comp}.

In the case of the W horizon, the deviations are always smaller than
0.2 degrees, and in most of the cases they are less than 0.1 degrees,
i.e. well within the expected rms errors. In general, the SRTM DEM
gives a better result than the ASTER one (Fig.~\ref{fig:comp}, left
panels). Things are different for the E horizon, for which the match
is good only along the smooth declining ridge of Mt. Faet in the
azimuth range 68$^\circ$--78$^\circ$, where the horizon is at a
distance of about 1.9 km (Fig.~\ref{fig:comp}, right panels). Between
64$^\circ$ and 68$^\circ$, where the distance drops to about 1.2 km,
the deviations exceed 0$^\circ$.5. But the largest discrepancies are
seen for azimuths larger than 84$^\circ$, where they are larger than
1$^\circ$.5. This very clearly illustrates the effect of a close
horizon coupled to a steep slope, where the smoothing intrinsic to the
DEM data produces systematically lower elevations and, in turn,
altitudes.

These two examples demonstrate that the procedure described here, when
used in conjunction to SRTM90 or ASTER DEMs, gives results accurate to
$\sim$0.1 degrees only for horizon distances larger than $\sim$10
km. This does not exclude that acceptable results may be achieved also
for shorter distances, but this will depend on the horizon morphology,
and cannot be assumed a priori.

\begin{figure}
\centering
\includegraphics[width=8cm]{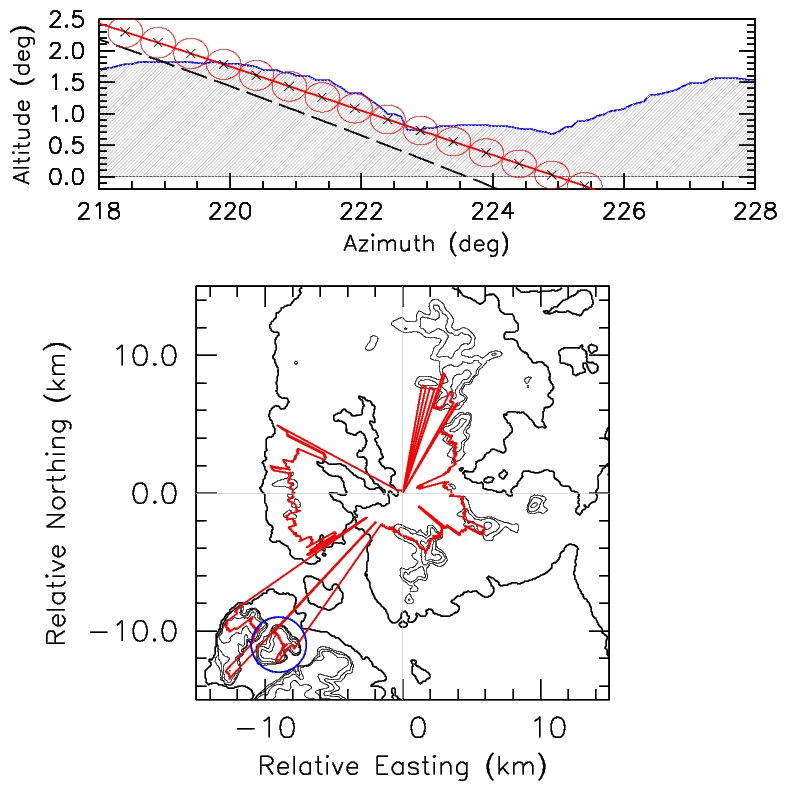}  
\caption{\label{fig:maeshowe}Lower panel: horizon location for
  Maes Howe. The map is a contour plot of the SRTM DEM. The position of
  Ward Hill is marked by a blue circle.  Upper panel: apparent path of
  the sun on December 1st ($\delta_\odot$=$-$21$^\circ$.9; solid red
  line). The dashed line traces the real path of the sun (see text). The red
  circles mark the position of the sun (to scale) with a time interval
  of 2 minutes. The blue curve is the SRTM synthetic horizon as seen from
  Maes Howe.}
\end{figure}

\section{\label{sec:example}Example application: the case of Maes Howe}

The method described in this paper can be applied any time there is
the need of having a handy description of the natural horizon in an
archaeo-astronomical analysis. To show the potential of the technique,
I close this article with an application to a famous and well studied
case, where the horizon shape is playing a fundamental role: Maes Howe.

Maes Howe is a neolithic site located on Mainland, Orkney, Scotland
(3$^\circ$.1879 W, 58$^\circ$.9981 N, 20 m a.s.l.). Besides being the
site of a chambered cairn and passage grave, Maes Howe offers a
spectacular event. Around twenty days before (and after) the winter
solstice, the sun sets below the horizon defined by Ward Hill, placed
at about 14 km SW of Maes Howe. However, after 7-8 minutes, it
reappears for a couple of minutes, just before definitely setting (Reijs
\cite{reijs98, reijs00}. See also Magli \cite{magli}, his Figs.~3 and
4). This is due to a combination of the sun apparent path and the
morphology of the horizon.

This phenomenon is very well reproduced using the synthetic horizon
computed for Maes Howe, once one applies a standard astronomical
refraction correction to the position of the sun\footnote{This
  correction is valid only on average, as the exact refraction depends
  on the physical conditions along the line of sight, and it is highly
  variable. See for instance Sampson et al.
  (\cite{sampson03}).}. This is illustrated in
Fig.~\ref{fig:maeshowe}, where I have traced the apparent trajectory
of the sun over the horizon. The sun reappears for a couple of minutes
at an azimuth around 223$^\circ$, in full agreement with direct
measurements (Reijs \cite{reijs98}), when the sun declination is
$\delta_\odot$=$-$21$^\circ$.9 (corresponding to December 1st of
today's calendar). An accurate reproduction is obtained also for the
sun reappearance in the direction of Kame of Hoy, which takes place
for $\delta_\odot=-$17$^\circ$.3 (February 1st), at an azimuth of
about 235$^\circ$.5. Since in 2700 BCE (the supposed date for the
construction of Maes Howe) the obliquity was about 23$^\circ$.8,
things were not very much different, and the same phenomenon took
place about 22 days before and after winter solstice. Having the
synthetic horizon, one can now ask what would be the azimuth of the
setting sun on the solstice. This turns out to be 216$^\circ$.8 (disk
center), and the horizon altitude is 1$^\circ$.3, but in this case no
reappearance is possible.  As minimum and maximum azimuths for which
the sun can shine on the back of the chamber are approximately
217$^\circ$ and 223$^\circ$ (Reijs \cite{reijs98}), both the solstice
setting and the sun reappearance should be observable from within the
chamber itself.

\begin{figure}
\centering
\includegraphics[width=8cm]{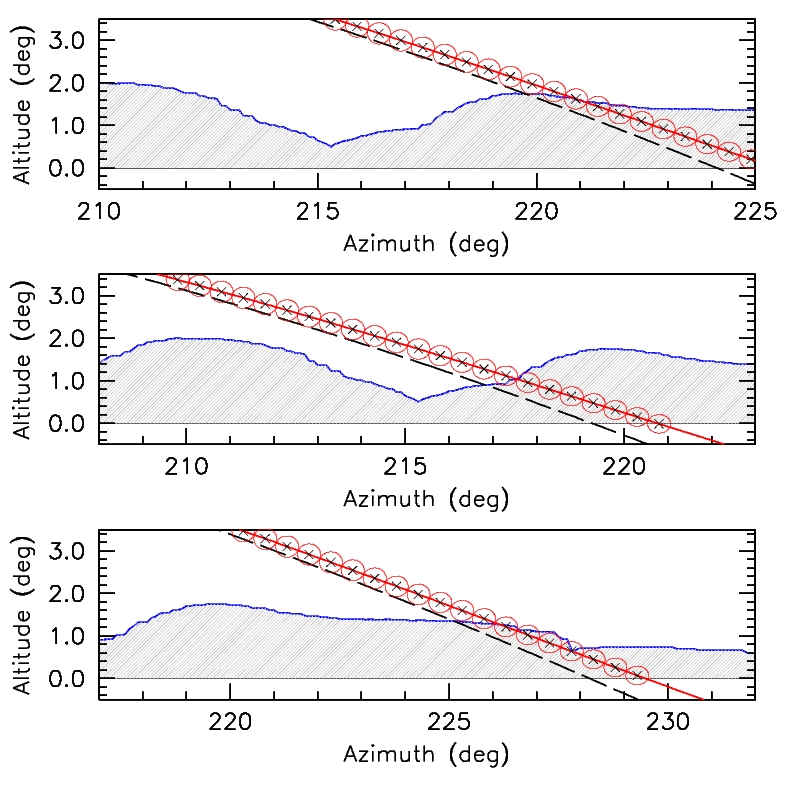}  
\caption{\label{fig:brodgar} Synthetic horizon computed for the Ring
  of Brodgar and apparent sun path for $\delta_\odot$=$-$21$^\circ$.7
  (top), $-$23$^\circ$.5 (middle), and $-$20$^\circ$.1 (bottom).}
\end{figure}

The procedure outlined in this paper could be used, for instance, to
test whether the reappearing sun is visible from other sites of
archaeological relevance close to Maes Howe. As an example I analyze
the case of the Ring of Brodgar\footnote{See {\tt
    http://www.orkneyjar.com/history/brodgar/}} (3$^\circ$.2280 W,
59$^\circ$.0014 N). This stone circle, located between the Lochs of
Stenness and Harray, has a diameter of about 104 m, and is the third
largest in the British Isles (Ruggles \cite{ruggles}). As the
calculations show, around $\pm$20 days from winter solstice the sun
sets over the Kame of Hoy, about 10$^\circ$ S of Ward Hill, without
re-emerging from behind it (Fig.~\ref{fig:brodgar}, top panel). At
present day's winter solstice, the sun still sets behind the Kame of
Hoy, without re-appearing (Fig.~\ref{fig:brodgar}, middle panel). Even
for the increased obliquity expected for 2700 BCE the apparent sun
path does intersect the profile of Ward Hill.  On the contrary, the
reappearance of the sun is in principle observable from the Ring of
Brodgar when $\delta_\odot$=$-$20$^\circ$.1 (November 22/January 21),
that is about one month before and after winter solstice
(Fig.~\ref{fig:brodgar}, bottom panel).

\section{\label{sec:concl} Conclusions}

The procedure I have outlined in this paper allows one to easily
compute a synthetic horizon profile for any site. The accuracy in the
final product depends on the resolution of the DEM and its vertical
accuracy. Rms deviations of $\sim$0$^\circ$.1 degrees are achievable
with the SRTM90 data when the horizon is at distances of the order of 10
km. For shorter distances, the deviations are expected to be larger,
especially if the horizon is defined by steep mountain profiles. In
those cases the morphology of the area surrounding the site of
interest needs to be examined more closely, and the synthetic horizon
validated with direct theodolite measurements obtained along some critical
directions.

The method can be applied during the exploratory phase of an
archaeoastronomical site's study, either to plan direct on-site
surveys or to test working hypotheses. When accuracies of a few tenths
of a degree are sufficient (as is certainly the case for solar
alignments), the results provided by the method can be used directly
for the orientation analysis, saving a significant amount of time
during the field work.

\acknowledgements This work has made use of the Shuttle Radar
Topographic Mission Data. ASTER GDEM is a product of METI and
NASA. All the results presented in this article were computed using C
programmes coded by the author. These can be made available to
  the reader upon request. The author is deeply indebted to V. Reijs
  for his very useful comments, and to the referee, D. Valls-Gabaud,
  for his critical review of the manuscript.

\newpage%%%%%%%%%%%%%%%%%%%%%%%%%%%%%%%%%%%%%%%%%%%%%%%%%%%%%%

\appendix

\section{\label{sec:app} Derivation of Earth's curvature correction}

In the following I will assume Earth can be described by an ellipsoid
with semi-axis $a$ and $b$. However, for the sake of simplicity, I
will also assume that locally it can be approximated by a sphere
having the curvature radius of the ellipsoid at the site under
examination. If $\phi$ is the site geodetic latitude, this is given by

\begin{equation}
\label{eq:radius}
R(\phi) = \frac{ab}{\sqrt{a^2 \sin^2 \phi + b^2\cos^2 \phi}}\;.
\end{equation}

\begin{figure}
\centering
\includegraphics[width=8cm]{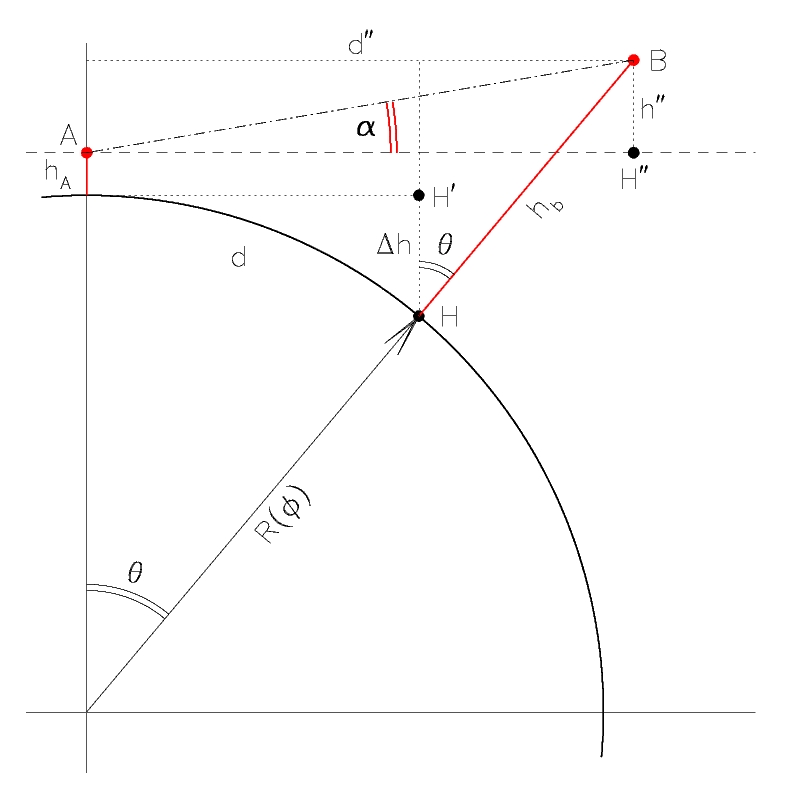}  
\caption{\label{fig:geom}Geometry of the problem.}
\end{figure}

Throughout this paper I adopt the WGS84 ellipsoid, which has
$a$=6378.1 km and flattening $f$=1.0/298.25722356 ($b$=6356.8 km). If
we now imagine to have an observer placed in $A$, at elevation $h_A$
above the ellipsoid, looking at a site $B$ placed at elevation $h_B$
and with an angular separation on the ellipsoid defined as $\theta$
(see Figure~\ref{fig:geom}), the altitude $\alpha$ of $B$ as seen from
$A$ above the local horizontal plane is given by

\begin{equation}
\alpha = \arctan \left ( \frac{h^{\prime\prime}}{d^{\prime\prime}} \right ) \;,
\end{equation}

\noindent where $d^{\prime\prime}$ and $h^{\prime\prime}$ are defined
as in Figure~\ref{fig:geom}. While $d^{\prime\prime}$ is the distance
between $A$ and $B$ projected on the local horizontal plane,
$h^{\prime\prime}$ is the elevation of $B$ above the same plane. The
whole problem reduces to determining these two lengths.

Looking at Figure~\ref{fig:geom} we immediately note that:

\begin{displaymath}
d^{\prime\prime}= \left ( R+h_B \right ) \;\sin \theta \;.
\end{displaymath}

If we define $\Delta h = HH^\prime$ (see Figure~\ref{fig:geom}), then
we have that:

\begin{displaymath}
h^{\prime\prime} = h_B\cos \theta - h_A - \Delta h\;.
\end{displaymath}

Since $\Delta h=R (1-\cos \theta)$, we finally obtain:

\begin{displaymath}
h^{\prime\prime}= h_B \cos \theta - h_A - R(1-\cos \theta).
\end{displaymath}

Therefore, the elevation of $B$ as seen from $A$ is given by the
following expression:

\begin{equation}
\label{eq:alpha}
\alpha = \arctan 
\left [  
\frac{h_B \cos \theta - h_A - R(1-\cos \theta)}{(R+h_B)\sin \theta}
\right ]\;.
\end{equation}

This formula can be approximated considering that in the practical
applications $\theta$ is going to be at most a few degrees. In fact,
if we consider two points separated by an ellipsoidal distance $d$=300
km, the angular separation is $\theta\approx d/R$=2$^\circ$.7. In
these circumstances, one can use the following approximate expressions
for the two trigonometric functions: $\cos \theta \approx
1-\frac{1}{2} \theta^2$ and $\sin \theta \approx \theta$ (where
$\theta$ is expressed in radians).

After substituting them in Equation~\ref{eq:alpha}, and considering
that in all practical cases $h_B/R\ll$1, we finally arrive at the
following expression:

\begin{equation}
\label{eq:approx}
\alpha \approx \arctan \left [
\frac{h_B-h_A}{d} - \frac{1}{2} \frac{d}{R}
\right ]\; ,
\end{equation}

\noindent which can be readily used to derive the curvature-corrected
altitude.

\begin{thebibliography}{}
\bibitem[1980]{bomsford} Bomsford, G., 1980, Geodesy (Oxford: Clarendon), 855
\bibitem[2007]{srtm} Farr, T.G., et al., 2007, Rev. Geophys., 45, RG2004
\bibitem[1989]{utm} Hager, J.W., Behensky, J.F. \& Drew, B.W., 1989, 
  DMA Technical Manual, DMATM 8358.2 (Fairfax, VA: Defense Mapping Agency)
\bibitem[2009]{magli} Magli, G., 2009, Mysteries and Discoveries of 
  Archaeoastronomy (Berlin: Springer Verlag), 54
\bibitem[1998]{reijs98} Reijs, V.M.M., 1998, 3rd Stone Magazine, 32, 18
\bibitem[2001]{reijs00} Reijs, V.M.M, 2001, The reappearing sun at Orkney, 
  {\tt http://www.iol.ie/$\sim$geniet/maeshowe/}
\bibitem[1999]{ruggles} Ruggles, C., 1999, Astronomy in Prehistoric 
  Britain and Ireland (New Haven: Yale University Press)
\bibitem[2003]{sampson03} Sampson, R.D., Lozowski, E.P., Peterson, A.E. \&
  Hube, D.P., 2003, Pub. Astron. Soc. Pac., 115, 1256
\bibitem[1975]{vincenty} Vincenty, T., 1975, Survey Review, 23, 88
\end{thebibliography}
\end{document}